\documentclass[final,twocolumn]{elsarticle}

\usepackage{graphicx}
\usepackage{bm}

\journal{Comp. Phys. Commun.}

\begin{document}

\def\Msun{M_{\odot} }

\begin{frontmatter}



\title{The collisions of high-velocity clouds with the galactic halo}


\author[1,2]{Petr Jel\'\i nek}

\address[1]{University of Vienna, Institute of Astronomy, T\"{u}rkenschanzstra{\ss}e 17,\\
1180 Vienna, Austria}
\address[2]{University of South Bohemia, Faculty of Science, Brani\v sovsk\'a 10,\\
370 05 \v{C}esk\'e Bud\v{e}jovice, Czech Republic}

\author[1]{Gerhard Hensler}

\begin{abstract}
Spiral galaxies are surrounded by a widely distributed hot coronal gas
and seem to be fed by infalling clouds of neutral hydrogen gas
with low metallicity and high velocities.
We numerically study plasma waves produced by the collisions of
these high-velocity clouds (HVCs) with the hot halo gas and with
the gaseous disk. In particular, we tackle two problems numerically:
1) collisions of HVCs with the galactic halo gas and
2) the dispersion relations to obtain the phase and group velocities
of plasma waves from the equations of plasma motion as well as
further important physical characteristics such as
magnetic tension force, gas pressure, etc.
The obtained results allow us to understand the nature of MHD waves
produced during the collisions in galactic media and lead to the
suggestion that these waves can heat the ambient halo gas.
These calculations are aiming at leading to a better understanding
of dynamics and interaction of HVCs with the galactic halo and of
the importance of MHD waves as a heating process of the halo gas.
\end{abstract}


\end{frontmatter}


\section{Introduction}
\label{}Spiral galaxies are surrounded by hot coronal gas for whose
existence three main origins are under debate: 1) Most plausibly,
star-formation activity in the galactic disk (stellar wind and
supernova type II explosions) pushes hot gas into the halo, by this,
producing the hot-gas halo \cite{Strickland,Tuellmann}. 2) Galaxy
growth within the $\Lambda$CDM cosmology will inherently lead to gas
accretion and by this produce a hot halo with virial temperatures if
the cooling time is too long as for temperatures above $10^6$ K and
thus for large galactic masses of $M > 10^{11} \Msun$. 3) And at
least, cosmological simulations suggest that 30 -- 40$\%$ of all
baryons reside in a cosmic web of shock-heated warm-to-hot
intergalactic medium accumulating as an extended hot halo around
massive galaxies \cite{Rasmussen}.

Another gas phase existing in the halo of spiral galaxies are
high-velocity clouds (HVCs). In the Milky Way these gas complexes are
detected in HI at high galactic latitudes \cite{Wakker} and are
falling with high speeds towards the galactic disk. Since
galactic halo gas is preferentially assumed to be produced through
the above-mentioned process 1), some of this hot gas is suggested
to condense, cool \cite{Collins}, and to fall back to the galactic
plane, or HVCs can stem directly from galactic HI gas that is
swept-up by superbubbles, lifted into the halo, and returns on
ballistic trajectories.
While the first is, however, contradicted by Binney and collaborators
\cite{Binney} because of heat conduction, the second process would
require also positive HI cloud velocities which are, however, not
observed.

This expected galactic fountain effect \cite{Shapiro} can, however,
not explain HVC velocities close or even above the galactic escape
value. Moreover, HVCs contain clearly less than solar metal
abundances and have probably larger distances
\cite{Wakker01,Wakker08} than achievable from superbubble expansion,
so that their origin from the galactic disk is very unlikely. In
addition, there also exists a population of intermediate velocity HI
clouds (IVCs). With their higher metallicity and far lower velocity
of only $50-100~\mathrm{km\cdot s^{-1}}$ these clouds as well as
those even slower (and smaller), so-called low-velocity clouds
(LVCs), are also likely to reflect at least partly back-falling
bullets expected from a galactic fountain.

A not insignificant part of LVCs, however, can be assumed to have
passed the galactic halo and is decelerated by the drag of the ram
pressure exerted by the hot halo gas \cite{Benjamin}. This also
leads to gas stripping from the clouds which is discernible as
head-tail structure of HVCs \cite{Bruens}. As observed in several
HVCs before and expected theoretically \cite{Wolfire}, the HVC
125+41-207 shows a two-phase structure with a low and high velocity
dispersion for column densities larger than 2$\times$ 10$^{15}$
m$^{-2}$.

Since in external galaxies seen edge-on like e.g. NGC 891 and NGC 4361
the hot halo gas is permeated with vertical magnetic field lines,
and because the galactic hot halo gas is assumed to be mostly replenished
by superbubbles which open perpendicular to the gaseous disk and, by
this, also open the galactic magnetic field perpendicularly as it is
coupled to the ionized gas, HVCs must be assumed to interact with
the halo magnetic field. Unfortunately, topology, strength, and
signatures of the galactic magnetic field are not well known and not yet
manifested by observations. Consequently, the magnetic coupling of
HVCs with the halo magnetic field is still vague and not yet elaborated.
This innocence is not only caused by uncertainty of the field
topology but mainly due to the total absence of magnetic field
measurements in HVCs.

Only Zimmer et al. \cite{Zimmer} interpreted observations of X-ray gas
connected with HVCs \cite{Kerp}, and preferentially with the
HVC complex C, as signatures of magnetic reconnection and explored
the magnetic interaction of HVCs with the Reynolds layer \cite{Reynolds}.
Another and far more realistic interpretation of the multi-gas phase
coincidence in HVCs is the sweep-up of hot halo gas by the large
complexes on their passage through the halo.
HVC models by Vieser \& Hensler \cite{Vieser} in which the clouds survive the disruption
by Kelvin-Helmholtz instability due to self-gravitation, heat conduction,
and cooling, the gas accumulation from such surrounding hot gas has
been demonstrated.

Subsequently, Santillan et al. \cite{Santillan} investigated the
collision of HVCs with the magnetic galactic gas disk by means
of 2D numerical models. In contrast to the model assumptions of
both studies no HVCs, i.e. large cloud complexes with velocities
of $-200~\mathrm{km\cdot s^{-1}}$ and below, are found close to the
galactic disk. Another limitation is that the magnetic field (in 2D)
is only directed parallelly to the galactic plane.

\section{Motivation of the numerical study}
One of the unsolved problems in solar plasma physics is the heating
of the solar corona. There exist mainly two possible explanations of
this interesting problem -- magnetic energy releasing and heating by
the reconnection of magnetic field and heating the solar corona by
plasma waves. In recent years, both of possible mechanisms were
solved by many authors theoretically and also numerically (see e.g.
\cite{Kliem}, \cite{Jelinek1} or \cite{Jelinek2}).

\begin{figure}[h!]
\hspace{-0.3cm} \vspace{-0.7cm}
\begin{center}
\includegraphics[width=3.2in]{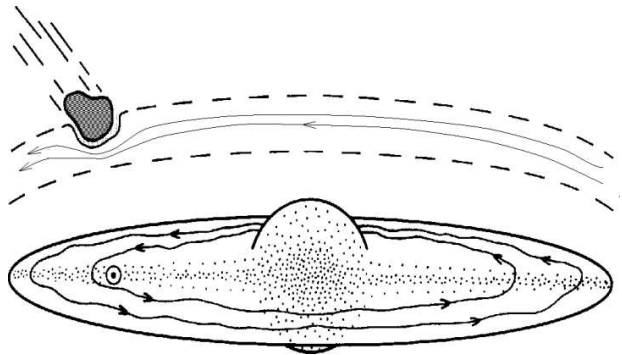}
\caption{A HVC collides with the Reynolds layer, an ionized hydrogen
layer of the galactic halo which includes magnetic fields. Picture
taken and redrawn from \cite{Zimmer}.} \label{fig1}
\end{center}
\end{figure}

Since superbubbles expanding into the halo are supposed to cool
adiabatically and the halo gas should continue at lower temperature
to cool by radiation and by cloud evaporation, a continuous hot gas
supply is necessary. Because of the complexity of the multi-phase
gas structure of the halo there is still a lack of sufficiently
detailed quantitative explorations and results about its
thermodynamics and magnetohydrodynamics \cite{Henley}. An additional
heating source would therefore elegantly provide a solution to
prevent the halo cooling. During the collisions of HVCs on
trajectories oblique or perpendicular to the magnetized galactic
halo or galactic disk, the magnetic field lines along the galactic
disc are distorted \cite{Zimmer}, gas is compressed, and MHD waves
are generated.

Such MHD waves (including reconnection of magnetic field) may, on
one hand, lead to heating of the ambient gas and can,
furthermore, influence the evolution and the dynamics of HVCs. This
process is analogous to the coronal heating in the sun \cite{Hirth}.
This similarity of MHD processes encourage us to begin analogous MHD
studies of HVCs passing through the halo and approaching the
galactic disk while interacting with the magnetic field.

\section{Numerical model}
\subsection{Simulations of collisions of HVCs with galactic halo}
In plasma physics, there exist several methods, how the plasma
dynamics can be described and calculated, see e.g. \cite{Jelinek3},
\cite{Simek}.

For the calculations of HVCs' dynamics we used the full set of MHD
equations \cite{Chung} with the addition of gravitational term:
\begin{equation}\label{eq1}
\frac{\partial \varrho}{\partial t}=-\nabla\cdot(\varrho \bm{v}),
\end{equation}
\begin{equation}\label{eq2}
\varrho \frac{\partial \bm{v}}{\partial t} +
\varrho(\bm{v}\cdot\nabla)\bm{v}=-\nabla
p+\frac{1}{\mu_0}(\nabla\times\bm{B})\times\bm{B} + \varrho \bm{g},
\end{equation}
\begin{equation}\label{eq3}
\frac{\partial \bm{B}}{\partial t}=\nabla\times(\bm{v}\times\bm{B}),
\end{equation}
\begin{equation}\label{eq4}
\frac{\partial U}{\partial t}=-\nabla\bm{S},
\end{equation}
\begin{equation}\label{eq5}
\nabla\cdot\bm{B}=0.
\end{equation}
Here $\varrho$ is a mass density, $\bm{v}$ (relative) flow velocity,
$p$ gas pressure, $\bm{B}$ is the magnetic field and $\bm{g}$ is the
gravitational acceleration calculated by means of Eq.
(\ref{eq17}).

The plasma energy density $U$ is given by:
\begin{equation}\label{eq6}
U=\frac{p}{\gamma - 1}+\frac{\varrho}{2}v^2+\frac{B^2}{2\mu_0},
\end{equation}
with the adiabatic coefficient $\gamma = 5/3$, and the flux vector
$\bm{S}$ is expressed as:
\begin{equation}\label{eq7}
\bm{S}=\left(U+p+\frac{B^2}{2\mu_0}\right)\cdot
\bm{v}-(\bm{v}\cdot\bm{B})\frac{\bm{B}}{\mu_0}.
\end{equation}

The magnetohydrodynamic equations (\ref{eq1})-(\ref{eq4}) were
transformed into a flux conserving form, i.e.:
\begin{equation}\label{eq8}
\frac{\partial \mathbf{\Psi}}{\partial t} + \frac{\partial
\mathbf{F}(\mathbf{\Psi)}}{\partial x} + \frac{\partial
\mathbf{G}(\mathbf{\Psi)}}{\partial y}= 0,
\end{equation}
and were solved numerically. The vector $\mathbf{\Psi}$ in our
two-dimensional case is expressed as:
\begin{equation}\label{eq9}
\mathbf{\Psi}=\left(\begin{array}{c} \rho \\ \rho v_x \\ \rho v_y \\
B_x \\ B_y \\ U~\end{array} \right).
\end{equation}
The vector functions $\mathbf{F} (\mathbf{\Psi})$ and $\mathbf{G}
(\mathbf{\Psi})$ are too complex to be presented here, for more
information see e.g. \cite{Chung}.

For the numerical solution of this type of equations the two-step
Lax--Wendroff algorithm is used. The numerical region is oriented in
the $x,y$-plane, implemented at $0 \leq x \leq L$ and $0 \leq y \leq
L$ and covered by a uniform grid with $200 \times 200$ cells. We
choose the $x$ coordinate to be oriented in the galactic plane,
while $y$ points vertically into the halo.

Open boundary conditions are applied and the time step satisfies
Courant-Friedrichs-Levy condition in the form \cite{Chung}:
\begin{equation}\label{eq10}
\Delta t \leq \frac{\mathrm{CFL}\cdot\Delta x}{\max(c_\mathrm{s} +
|\mathbf{v}|)},
\end{equation}
where the Courant number CFL is set to 0.8.

\subsection{Calculations of phase and group speeds}
For the calculations of phase and group velocities of plasma waves
in the galactic gas, we numerically solve the wave equation for
plasma motions, where the equilibrium parameters (density and
pressure) depend on $y$ (e.g. \cite{Smith}):
\begin{equation}\label{eq11}
\frac{\mathrm{d}}{\mathrm{d}y}\left[f(y)\frac{\mathrm{d}v_y}{\mathrm{d}y}\right]+\varrho(\omega^2
- k_x v_{\mathrm{Alf}}^2)v_y=0.
\end{equation}
Here $v_y$ is the velocity component normal to the magnetic field,
$\omega$ is the frequency, $k_x$ is the longitudinal wavenumber
along the $x$-axis, and Alfv\'en speed $v_{\mathrm{Alf}}^2 = B^2 /
\mu_0 \varrho$.

The velocity $v_x$, parallel to the magnetic field is given by:
\begin{equation}\label{eq12}
v_x = -\frac{i k_x c_\mathrm{s}^2}{(\omega^2 - k_x^2
c_\mathrm{s}^2)} \cdot \frac{\mathrm{d}v_y}{\mathrm{d}y},
\end{equation}
where $c_\mathrm{s} = (\gamma p / \varrho)^{1/2}$ is the sound
speed.

The function $f(y)$ from Eq. (\ref{eq11}) is expressed as:
\begin{equation}\label{eq13}
f(y) = \frac{\varrho c_\mathrm{f}^2(\omega^2 - k_x^2
c_\mathrm{T}^2)}{(\omega^2 - k_x^2 c_\mathrm{s}^2)}.
\end{equation}
The tube speed $c_\mathrm{T}$ and fast speed $c_\mathrm{f}$ are
implied as $c_\mathrm{T} = c_\mathrm{s} v_{\mathrm{Alf}} /
(c_\mathrm{s}^2 + v_{\mathrm{Alf}}^2)^{1/2}$ and $c_\mathrm{f} =
(c_\mathrm{s}^2 + v_{\mathrm{Alf}}^2)^{1/2}$, respectively.

Eq. (\ref{eq11}) has a  singular point which is called the cusp
resonance or cusp singularity. This point plays an important role in a
case of slow magnetoacoustic waves, whereas it is not seen in
numerical simulations of the fast magnetoacoustic waves (e.g.
\cite{Zhukov}).

The second-order ordinary differential equation is rewritten in
terms of two first-order equations of the new functions $\xi_1$ and
$\xi_2$:
\begin{equation}\label{eq14}
\xi_1 = f(y)\frac{\mathrm{d}v_y}{\mathrm{d}y}, \hspace{1cm} \xi_2 =
v_y,
\end{equation}
such that:
\begin{equation}\label{eq15}
\frac{\mathrm{d}\xi_1}{\mathrm{d}y}=\varrho(k_x
v_{\mathrm{Alf}}^2-\omega^2 )\xi_2, \hspace{0.5cm}
\frac{\mathrm{d}\xi_2}{\mathrm{d}y}=\frac{\xi_1}{f(y)}.
\end{equation}
The boundary conditions at the point $y = 0$ for the ``kink" mode
are given by $\xi_1 = 0, \xi_2 = c$ and the ``sausage" mode
satisfies the boundary conditions $\xi_1 = c \cdot f(0), \xi_2 = 0$,
whereas the constant $c$ is arbitrary.

To obtain a solution of Eq. (\ref{eq11}) a fixed value of $k_x$
is used, integrating between $y = 0$ and $y=y_\mathrm{max}$ the two
first-order differential equations (\ref{eq14}) and (\ref{eq15}) by
means of the fourth-order Runge-Kutta method. Exact value of
the frequency $\omega$ is obtained by the bisection iteration method
when the velocity $v_y$ satisfies the boundary condition at the
second point $v_y(y=y_\mathrm{max}) = 0$ for both wave modes
(``kink" and ``sausage" mode).

The total gas pressure $p_\mathrm{tot} $(gas and magnetic) is expressed
as:
\begin{equation}\label{eq16}
\frac{i \omega}{\varrho}p_\mathrm{tot} =
-\frac{c_\mathrm{f}^2(\omega^2-k_x^2
c_\mathrm{T}^2)}{(\omega^2-k_x^2
c_\mathrm{s}^2)}\frac{\mathrm{d}v_y}{\mathrm{d}y},
\end{equation}
and for the calculation of the magnetic tension force $\mathbf{T}_1$
we use the equation in the form, see \cite{Smith}:
\begin{equation}\label{eq17}
\frac{i \omega}{\varrho}\mathbf{T}_1 = (\omega^2-k_x^2
c_\mathrm{s}^2)\frac{v_\mathrm{Alf}^2}{c_\mathrm{s}^2}\mathbf{\hat{e}_x}
- v_\mathrm{Alf}^2 k_x^2 v_y \mathbf{\hat{e}_y}.
\end{equation}
This force is a part of the Lorentz force $\mathbf{j} \times
\mathbf{B}$ and appears whenever the magnetic field lines are
curved. Magnetic tension force is determined by how much the
magnetic pressure changes with distance (for detailed information
see \cite{Smith}).

\section{Initial conditions}
\subsection{Collisions of HVCs with galactic halo}
Numerical solutions of MHD equations are performed in 2D space.
The absolute size of the simulation box amounts
$3~\mathrm{kpc}\times 3~\mathrm{kpc}$. The initial position of
modeled HVC is located $1250~\mathrm{pc}$ above the galactic plane
with an initial velocity of $v_0 = -200~\mathrm{km \cdot s^{-1}}$
\cite{Santillan}.

The mass density distribution of the galactic gas is given by the
equation:
\begin{eqnarray}\label{eq18}
\varrho(y) & = & \varrho_0 [0.6e^{-y^2/(280 \mathrm{pc})^2} + 0.37e^{-y^2/(540 \mathrm{pc})^2} +{}
\nonumber\\
& + & 0.1e^{-|y|/400 \mathrm{pc}}+0.03e^{-|y|/900 \mathrm{pc}}],
\end{eqnarray}
and for the expression of the gravitational acceleration we use the
equation \cite{Santillan}:
\begin{eqnarray}\label{eq19}
g(y) & = & 8\cdot10^{-7}[1 - 0.52e^{-|y|/325 \mathrm{pc}} -{}
\nonumber\\
& - &0.48e^{-|y|/900 \mathrm{pc}}] \: \mathrm{m}\cdot \mathrm{s}^{-2}.
\end{eqnarray}
The galactic midplane gas density in Eq. (\ref{eq18}) is $\varrho_0
= 2.24\times 10^{-21}~\mathrm{kg \cdot \mathrm{m}^{-3}}$ and the
initial magnetic field $B_0 = 5.0~\mathrm{\mu G}$. The total
pressure is given by the relation $p(y) = \int \varrho g
\mathrm{d}y$, with the boundary condition that $p_\mathrm{out} = p(y
= 5.0~\mathrm{kpc}) = 0~\mathrm{Pa}$. The galactic midplane value of
the total pressure $p_0 = 2.7 \times 10^{-13}~\mathrm{Pa}$
\cite{Santillan}.

\subsection{Phase and group speeds}
For both studied cases (uniform magnetic field parallel to the
galactic plane and galactic current sheet configuration), we assume,
in the state of equilibrium, initial plasma velocity $\bm{v} = 0$.

Equilibrium demands that the pressure (plasma plus magnetic) is uniform, i.e.:
\begin{equation}\label{eq20}
p+\frac{B^{2}}{2\mu_{0}}=\mathrm{const.}
\end{equation}

\subsubsection{Uniform magnetic field}
In this case, the magnetic field is assumed $\mathbf{B} = B_0 =
\mathrm{const.}$ in the whole simulation region. The total pressure
$p$ according to the Eq. (\ref{eq20}) is also assumed to be a
constant, $p = p_0$.

\subsubsection{Current sheet}
In the case of current sheet configuration the magnetic field is
given by the equation \cite{Kliem}:
\begin{equation}\label{eq21}
\mathbf{B} = B_0 \tanh\left[\frac{(y -
L/2)}{a}\right]\mathbf{\hat{e}_x},
\end{equation}
and the equilibrium in Eq. (\ref{eq18}) yields a plasma pressure given by:
\begin{equation}\label{eq22}
p(y) = p_0\, \mathrm{sech}^{2}\left[\frac{(y - L/2)}{a}\right],
\end{equation}
given that $p \rightarrow 0$ (cold plasma) as $(|y|/a)
\rightarrow\infty$ \cite{Smith}, and where $a = 0.5~\mathrm{kpc}$ is
the semi-width of the current sheet.

Numerical solutions of Eq. (\ref{eq11}) are performed only for the
so-called ``kink" mode \cite{Smith}, which well corresponds to the
situation of the HVC collision with galactic disk.

\section{Results and discussions}
In this section we present some selected numerical results obtained
by means of our simulations.

\subsection{Collisions of HVCs with galactic halo}
In Figs. 2 and 3 we present the ``time evolution" of a HVC collision
with galactic media. Fig. 2 shows the perpendicular collision,
whereas in Fig. 3 an oblique passage of the HVC with
$\alpha=45^{\circ}$ through the galactic halo is depicted.

The situations (a) show the gas density distribution at the
``beginning", i.e. $T_{a} = 0.6~\mathrm{Myr}$ after the onset of the
calculation, while the locations (b) present the density
distributions for the time $T_{b} = 7.9~\mathrm{Myr}$ in the case of
perpendicular collision (Fig. 2) and $T_{b} = 12.8~\mathrm{Myr}$ for
oblique cloud infall (Fig. 3).

\begin{figure}[h!]
\hspace{-0.5cm}
\begin{center}
\includegraphics[width=2.8in]{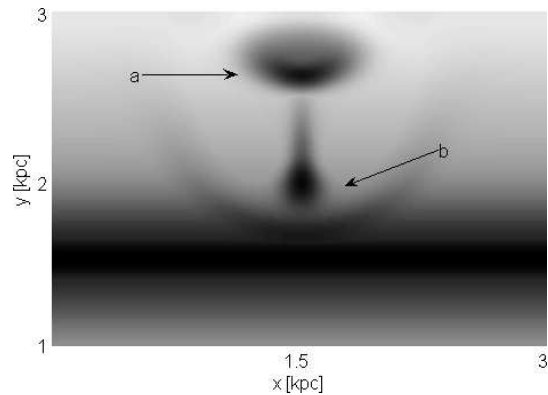}
\caption{The ``time evolution" of the collision of HVCs with
galactic halo in the case of perpendicular cloud infall. a: cloud at
0.6 Myr after the start of the models, b: position and structure at
time $T_{b}$ (see text).} \label{fig2}
\end{center}
\end{figure}

\begin{figure}[h!]
\hspace{-0.5cm}
\begin{center}
\includegraphics[width=2.8in]{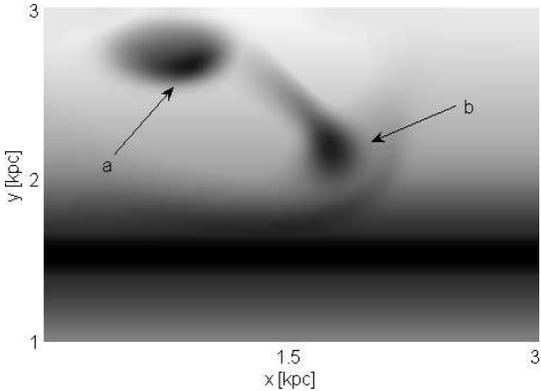}
\caption{The ``time evolution" of the collision of HVCs with
galactic halo in the case of oblique cloud infall under
$\alpha=45^{\circ}$. a: cloud at 0.6 Myr after the start of the
models, b: position and structure at time $T_{b}$ (see text).}
\label{fig3}
\end{center}
\end{figure}


In both cases at time $T_{b}$ a dense regime in front of the HVC is
clearly visible, where the galactic gas is compressed as the HVC
moves towards denser galactic disk and is decelerated.

Because this infall motion towards denser gas causes a bow shock to
form, consequently, also plasma waves are created.

In Figs. 4 and 5 are shown the total pressure of the galactic
gas and magnetic field lines for both studied cases, i.e.
perpendicular and oblique infall under angle $\alpha=45^{\circ}$.
The pressure on the head of the cloud is higher because the cloud
moves through gas with increasing density near galactic disk, and the
magnetic field lines are distorted because the magnetic field is
``frozen" to the HVC.

\begin{figure}[h!]
\hspace{-0.5cm}
\begin{center}
\includegraphics[width=2.8in]{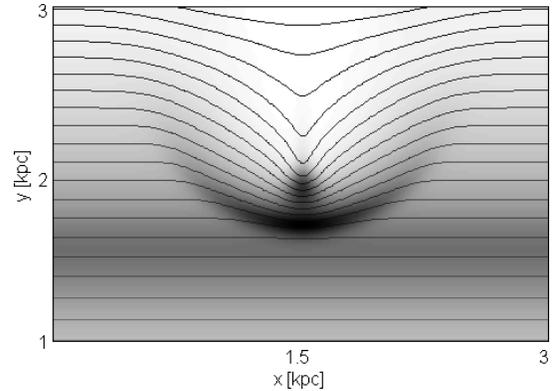}
\caption{The total pressure in the numerical region and magnetic
field lines distorted by the HVC for the perpendicular infall of the
cloud to the galactic disk at time $T_{b}$ (see text).} \label{fig4}
\end{center}
\end{figure}

\begin{figure}[h!]
\hspace{-0.5cm}
\begin{center}
\includegraphics[width=2.8in]{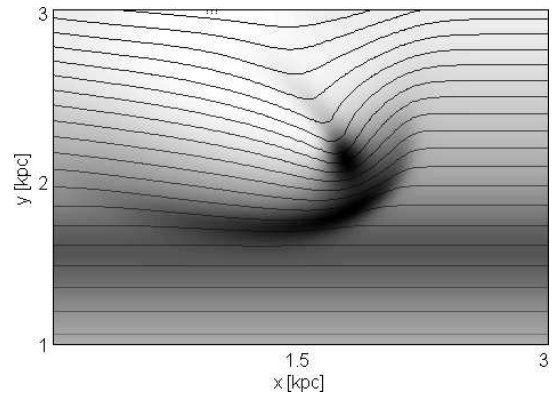}
\caption{The total pressure in the numerical region and magnetic
field lines distorted by the HVC for the oblique infall
($\alpha=45^{\circ}$) of the cloud to the galactic disk at time
$T_{b}$ (see text).} \label{fig5}
\end{center}
\end{figure}

\subsection{Phase and group speeds}
In Figs. 6 and 7 we present the phase and group speeds of wave
signal, plasma velocity, total gas pressure and magnetic tension
force for magnetic field oriented parallelly to the galactic plane
and for a current sheet configuration, respectively.

\begin{figure}[h!]
\hspace{-0.3cm}
\includegraphics[width=3.5in, height=2.5in]{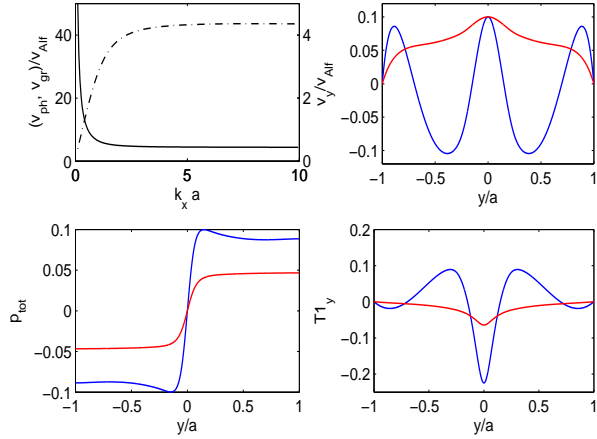}
\caption{Phase (solid) and group (dash-dotted) velocity of wave
signal (upper left); plasma velocity component $v_y$ (upper right),
total gas pressure (bottom left) and magnetic tension force (bottom right) for magnetic field oriented parallelly to the galactic
plane, wavenumbers $k_x a = 5$ (red) and $k_x a = 10$ (blue).}
\label{fig6}
\end{figure}

\begin{figure}[h!]
\vspace{-0.3cm}
\hspace{-0.3cm}
\includegraphics[width=3.5in, height=2.5in]{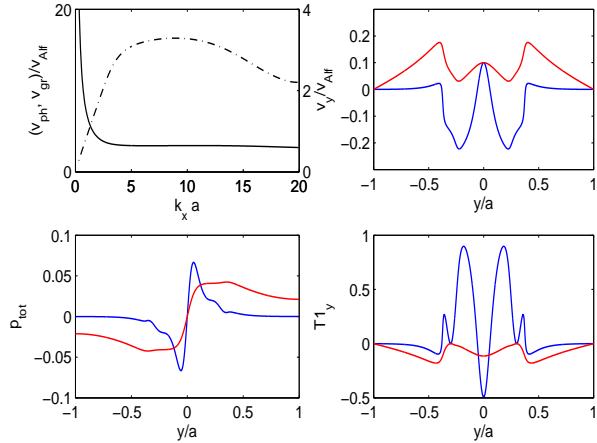}
\caption{Phase (solid) and group (dash-dotted) velocity of wave
signal (upper left); plasma velocity component $v_y$ (upper right),
total gas pressure (bottom left) and magnetic tension force (bottom right) for the current sheet configuration, wavenumbers $k_x a = 5$
(red) and $k_x a = 10$ (blue).} \label{fig7}
\end{figure}

Notice, that in the graph of the phase and group speeds the scale on the
left-hand side corresponds to the phase speed (solid line), whereas the scale
on the right-hand side corresponds to the group speed (dash-dotted line) of
the plasma waves. If we compare the phase and group speeds for both studied
cases, we can see that the profiles of phase speeds are similar, whereas the
group speeds have different shapes. In the case of the current sheet the group
speed becomes constant for high wavenumbers.

From the figures of total pressure $p_{\mathrm{tot}}$ we can discern
that for $k_x a = 5$ (red line) the lines are also the same, whereas
for $k_x a = 10$ (blue line) the total pressure is zero outside of
current sheet, because of constant (zero) plasma velocity $v_y$, see
Eq. (\ref{eq16}).

For the plasma velocity component $v_y$ in Fig. 7 it is clearly
visible, where the edge of the current sheet is located. Outside of
the current sheet the plasma velocity is zero for $k_x a = 10$ and
falls to zero in case of $k_x a = 5$. This means that practically
only the part of the galactic gas, where the magnetic field is not
constant, is moving. In the case of a parallel magnetic field,
galactic gas moves over the whole simulation region. It is probably
caused by the fact, that inside the current sheet is present the
Lorentz force $\bm{\mathrm{j}} \times \bm{\mathrm{B}}$ which does
not allow the plasma waves generated in the center of the current
sheet leave freely this space.

As one can see, the $y$-component of the magnetic tension force
$\mathbf{T}_{1}$ is in ``anti-phase" to the plasma $y$-velocity
component in all cases.

\section{Conclusions}
Our main aim is to investigate the effects of the collisions of HVCs
with both the galactic media, hot halo gas and magnetic field, on
their descendence towards the galactic disk. For this purpose here
we present a first insight with respect to two aspects by 2D
numerical models: at first, we solve numerically the set of MHD
equations and simulate the collisions of HVCs with the galactic
media for two cases (oblique and perpendicular collision). In the
second part of this paper, the equation of plasma motion is
treated numerically in order to obtain the dispersion relations of
waves in galactic media.

In the literature \cite{Zimmer} it is envisaged that the
reconnection of magnetic field lines in the galactic halo can lead
to its gas heating. Unfortunately, but as a physically reasonable
by-product, heating of the halo gas by plasma waves is not yet taken
into account. There exists practically no paper dealing with this
interesting problem.

Our first results demonstrate that plasma waves provide an
important heating process of the galactic halo gas in analogy to the
solar corona. From the above-mentioned reasons our results are new
and their further exploration of crucial importance for our
understanding of the heating processes in the interstellar medium in
general. Nevertheless, 2D simulations can only deal as a first
insight but 3D simulations are urgent for the MHD treatment to
deliver more accurate and reliable results. Therefore, as the next
step we will advance our models to 3D and to higher spatial
resolution. For this purpose we intend to use as a powerful, well
tested and widely applied tool for astrophysical simulations of many
applications the FLASH code \cite{Fryxell}, parallelized and with
adaptive mesh refinement. Recently, \cite{Kwak} performed first 3D
models using FLASH, but with the main emphasis on the
structural evolution of fountain clouds starting at a height of 5
kpc above the galactic disk. From the extension of our numerical
models we expect important results and information about the MHD
waves in galactic media. As further steps, also small-scale plasma
processes like heat conduction must be included, that leads to the
stabilization of clouds \cite{Vieser} by suppression of KH
instability.

\section*{Acknowledgements}
This research has been funded by the University of Vienna, partly
within the priority programme ``Computer-aided Sciences" under grant
No. FS538001.

P.J. acknowledges partial financial support of the
grant P209/10/1680 of the Grant Agency of the Czech Republic.

The authors thank the anonymous referee for valuable comments
which improve the quality of the paper.





\bibliographystyle{elsarticle-num}
\bibliography{<your-bib-database>}



\end{document}